\documentclass[10pt,twocolumn,a4paper]{article}
\usepackage{graphicx}
\usepackage{amsmath}
\usepackage{amssymb}
\usepackage{bm}
\usepackage{dcolumn}
\usepackage{subfigure}
\usepackage{units}
\usepackage[numbers]{natbib}

\input epsf
\newcommand{\BN}{\textit{h}-BN}

\begin{document}
\author{Menno Bokdam, Petr A. Khomyakov, Geert Brocks, Zhicheng Zhong and Paul J. Kelly}



\title{Electrostatic doping of graphene through ultrathin hexagonal boron nitride films}

\maketitle


The preparation of monolayers of graphite (graphene) has led to exciting discoveries associated with the unique electronic structure of this two-dimensional system.\cite{Novoselov:sc04,Geim:natm07} Transport experiments are usually performed in a field-effect device geometry with a graphene flake separated from a gate electrode by a dielectric spacer.\cite{Novoselov:nat05,Zhang:nat05} Doped Si is commonly used as the gate material and SiO$_2$ as the dielectric. Substrate inhomogeneities \cite{Ishigami:nanol07,Stolyarova:pnas07} and trapped charges \cite{Chen:natp08,Zhang:natp09} lead to local variations of the electrostatic potential at the SiO$_2$ surface which results in uncontrolled local electron- and hole-doping of the graphene sheet, so-called electron-hole puddles.\cite{Martin:natp08,Zhang:natp09}   

With a layered graphene-like honeycomb structure and a lattice constant only 1.7\% larger, insulating \cite{Watanabe:natm04} \BN{} would be the ideal substrate for graphene based devices;\cite{Giovannetti:prb07} the interaction between \BN{} and graphene sheets is so weak that even if their lattices are forced to be commensurate, the characteristic electronic structure of graphene is barely altered.  For an incommensurate stacking of graphene on \BN{}, the perturbation is even smaller, leading to an extremely well-ordered graphene layer with a very high carrier mobility.\cite{Dean:natn10,Xue:natm11,Decker:nanol11}

Like graphene, \BN{} layers can be prepared by mechanical exfoliation.\cite{Novoselov:pnas05} Cleaved layers can also be thinned to a single layer with a high-energy electron beam.\cite{Meyer:nanol09} Alternatively, \BN{} layers can be grown by chemical vapor deposition (CVD) on transition metals such as Cu or Ni, using precursors such as borazine (B$_3$N$_3$H$_6$) or ammonia borane (NH$_3$-BH$_3$).\cite{Song:nanol10,Shi:nanol10} With a proper choice of growth conditions, homogeneous ultrathin \BN{} only 1-5 atomic layers thick can be grown. Moreover, graphene can be grown by CVD on top of a \BN{} layer adsorbed on a metal substrate,\cite{Usachov:prb10} which is ideal for field-effect devices.

A field effect is created by applying a voltage difference between graphene and the metal substrate, resulting in charge accumulation or depletion in the graphene layer. We study this so-called electrostatic doping as a function of the thickness of the \BN{} layer and the applied voltage using first-principles (DFT) calculations for a Cu(111)$|$\BN{}$|$graphene structure. In the absence of an applied voltage, a spontaneous charge transfer across the \BN{} layer occurs between the metal substrate and the graphene, leading to an intrinsic doping of graphene. This transfer, which is driven by the difference in graphene and metal work functions, is strongly modified by the charge displacements resulting from the weak chemical interactions at the metal$|$\BN{} and at the \BN{}$|$graphene interfaces.

By varying the applied voltage, the doping level can be controlled. Experimentally the position of the Fermi level in graphene has a square-root like dependence on the voltage.\cite{Zhang:natp08,Zhang:natp09,Yu:nanol09,Xue:natm11,Decker:nanol11} We will demonstrate that this behavior is reproduced by our DFT calculations on Cu(111)$|$\BN{}$|$graphene and develop an analytical model that describes how the Fermi level depends on the applied voltage and the \BN{} layer thickness. 

\textit{Computational Details.} We use DFT at the level of the local density approximation (LDA), within the framework of the plane-wave PAW pseudopotential method,\cite{Blochl:prb94b} as implemented in VASP.\cite{Kresse:prb93,Kresse:prb96,Kresse:prb99} In our previous work we  found that the LDA gives a reasonable description of the geometries resulting from the weak interactions between \BN{} layers and between a \BN{} and a graphene layer.\cite{Giovannetti:prb07} The change in the electronic structure of such systems in an external field is also described well.\cite{Slawinska:prb10a,Slawinska:prb10b,Ramasubramaniam:nanol11} The LDA also describes the interaction between graphene and metal (111) surfaces very reasonably.\cite{Karpan:prl07,Giovannetti:prl08,Khomyakov:prb09,Wintterlin:ss09,Feibelman:prb08} We expect it to provide a description of the interaction between \BN{} and metal (111) surfaces of similar quality. In contrast, we found that PW91 or PBE GGA functionals (incorrectly) predict essentially no interlayer binding between \BN{} or graphene layers, or between metal (111) surfaces and \BN{} or graphene sheets. 

The Cu(111)$|$\BN{}$|$graphene structures are modeled in a supercell periodic in the $z$ direction with six Cu atomic layers in the (111) surface orientation, a slab 1-6 layers thick of \BN{}, a graphene monolayer, and a vacuum region of $\sim 15$ \AA. A dipole correction is applied to avoid spurious interactions between periodic images of the slab. We choose the lattice constant of graphene equal to its optimized LDA value $a = 2.445$ \AA, and scale the in-plane lattice constants of Cu(111) and \BN{} so that the structure can be represented in a $1\times 1$ graphene surface unit cell.\cite{Giovannetti:prb07,Karpan:prl07,Giovannetti:prl08,Khomyakov:prb09} The positions of the atoms in graphene and \BN{}, and of the top two atomic layers of the metal surface are allowed to relax during geometry optimization. The effect of the strain in the Cu and \BN{} structures on the electronic structure is very small. For instance, the work function of Cu(111) is increased by a mere 0.04 eV and the density of states at the Fermi level is unaltered. An electric field applied across the slab is modeled with a sawtooth potential.\cite{Resta:prb86}

In the most stable configuration of a \BN{} monolayer on Cu(111), the nitrogen atoms are adsorbed on top of Cu(111) surface atoms and the boron atoms occupy hollow sites. The calculated equilibrium separation between \BN{} and the Cu surface is 3.11 \AA. The graphene and \BN{} layers are stacked as in Ref.~\citenum{Giovannetti:prb07} with a calculated graphene-\BN{} equilibrium separation of 3.21 \AA, and a separation between \BN{} layers of $3.24$ \AA. The equilibrium structures are changed negligibly by the applied electric field, at least for field strengths up to $0.5$ V/\AA. The interaction of graphene with \BN{} is so weak that the characteristic linear band structure of graphene about the conical points is essentially preserved; a small band gap of $\sim 50$ meV is induced in graphene if the \BN{} and the graphene lattices are commensurate. If they are incommensurate, the gap seems to disappear.\cite{Xue:natm11,Dean:natn10,Decker:nanol11} In the following this point will not be important. 

\begin{figure} 
\includegraphics[scale=0.33, bb= 60 21 721 520]{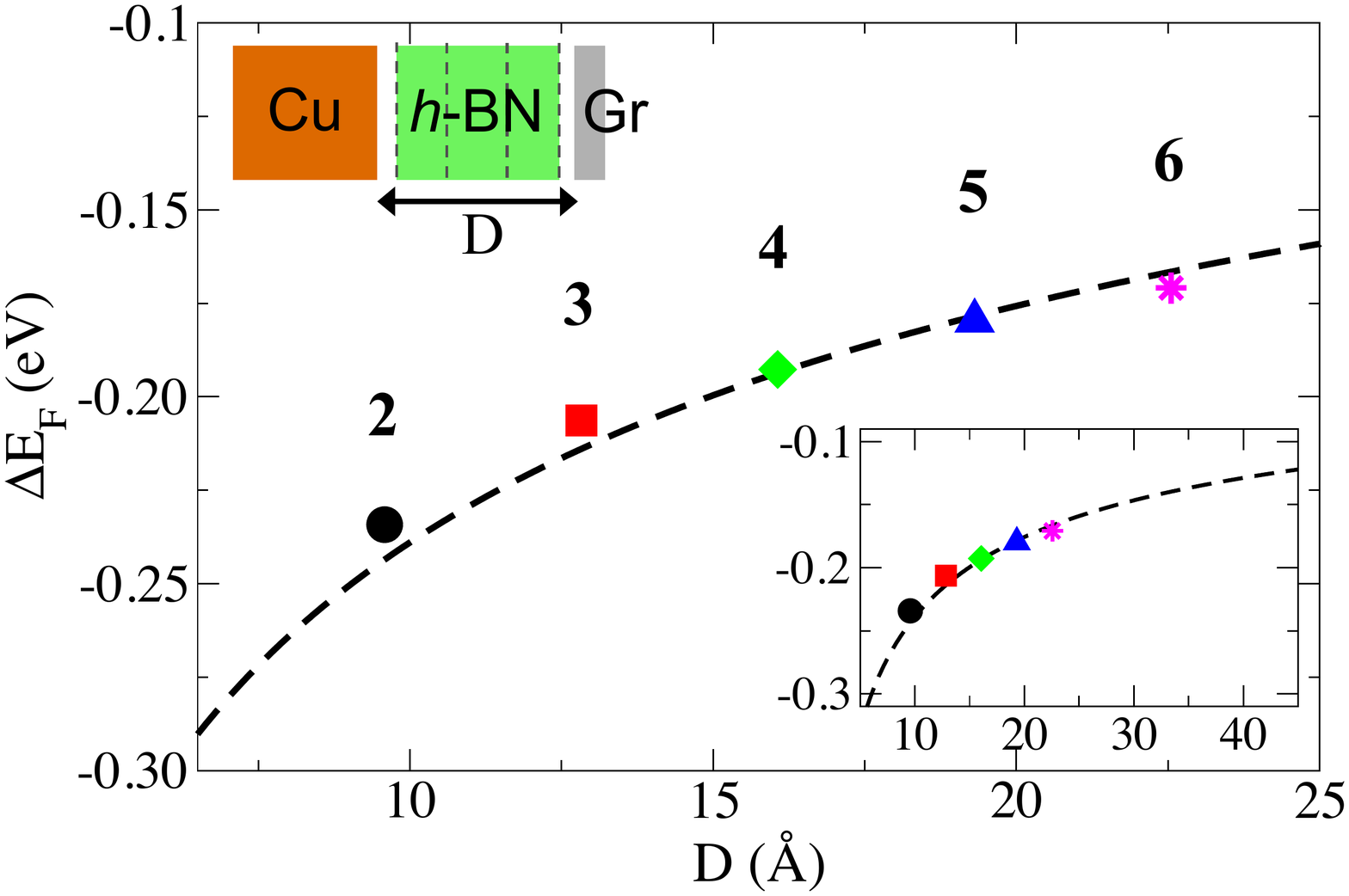}
\caption{The Fermi level shift $\Delta E_{\rm F}$ in graphene versus the thickness D of the \BN{} film. The symbols represent DFT results for Cu(111)$|$\BN{}$|$graphene structures with 2 - 6 layers of \BN{}. The dashed line represents the model of eq. \ref{fermishiftE}. } \label{A}
\end{figure}

\textit{Intrinsic Doping.} We monitor the doping of graphene in the Cu(111)$|$\BN{}$|$graphene structure by calculating the change of the Fermi level $\Delta E_{\rm F}$ with respect to neutral graphene. A computationally convenient way \cite{Giovannetti:prl08,Khomyakov:prb09} of characterizing $\Delta E_{\rm F}$ is in terms of the difference between the work functions $W$ of the metal$|$dielectric$|$graphene stack and $W_{\rm G}$ for free-standing graphene,  
\begin{equation}
\Delta E_{\rm F} = W - W_{\rm G}, \label{eq1}
\end{equation}
Negative (positive) values of $\Delta E_{\rm F}$ then correspond to $n$-type ($p$-type) doping. Figure \ref{A} shows $\Delta E_{\rm F}$  as a function of the number of \BN{} layers. These numbers are calculated without an external field which means that in the Cu(111)$|$\BN{}$|$graphene structure the graphene sheet is intrinsically doped with electrons. $\Delta E_{\rm F}$ decreases with the number of \BN{} layers. We will model this thickness dependence below.

The intrinsic doping originates from electrons that are transferred from the Cu electrode to the graphene sheet across the \BN{} layer. However, the calculated work functions of Cu(111) and graphene are $W_\mathrm{M}=5.25$ eV and $W_\mathrm{G}=4.48$ eV, respectively. If charge transfer is driven by the difference between the Cu substrate and graphene work functions alone, then establishing a common Fermi level would require transferring electrons from graphene to Cu and the result would be $p$-type doping, i.e. a positive value of $\Delta E_{\rm F}$, at variance with the results shown in figure \ref{A}. The \BN{} layer must therefore play a non-trivial role.

\begin{figure}
\includegraphics[scale=0.33, bb= 60 21 721 520]{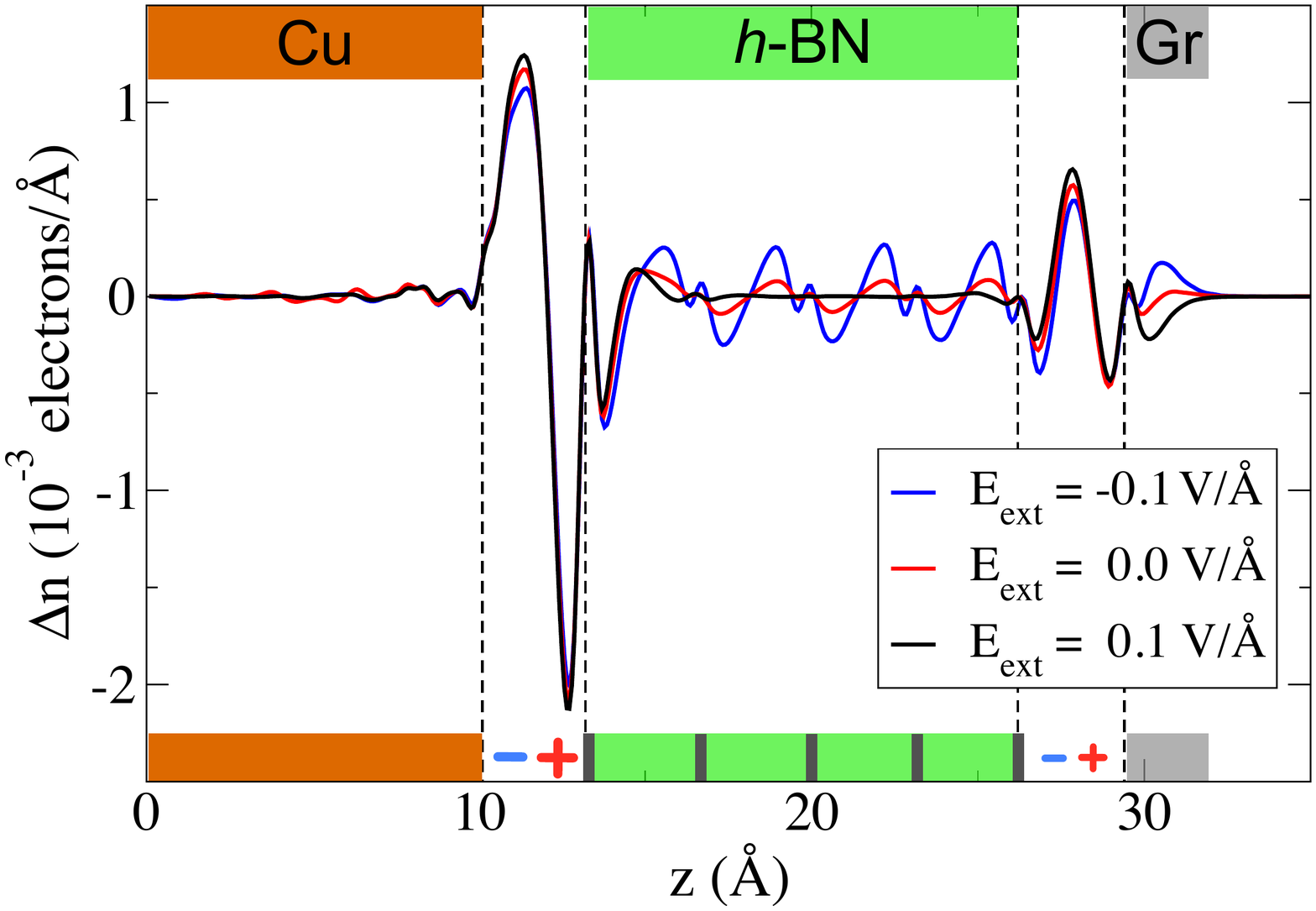}
\caption{
Plane-integrated electron density difference $\Delta n(z)$ for a Cu(111)$|$\BN{}$|$graphene structure with five \BN{} layers. The positions of the top layer of Cu atoms, the bottom and top \BN{} layers, and of the graphene layer are indicated by vertical dashed lines. Each of the three $\Delta n(z)$ curves represents a different value of the external electric field $E_{\rm ext}$.}
\label{fig-p3}
\end{figure}

The interaction between two materials at an interface generally results in the formation of an interface dipole. The latter can be visualized in terms of the electron rearrangement at the interface characterized by the electron density of the entire system minus the electron densities of the two separate materials (using identical, frozen atomic structures). As only the dependence perpendicular to the interface is relevant, it is convenient to work with plane-integrated electron densities $n(z) = \iint n(x,y,z) \, dxdy$ where the integration is over the surface unit cell. The electron displacement in a Cu(111)$|$\BN{}$|$graphene stack is then described by $\Delta n(z)=n_{\rm M|BN|Gr}(z)-n_{\rm M}(z)-n_{\rm BN}(z)-n_{\rm Gr}(z)$. The result for a Cu(111)$|$\BN{}$|$graphene structure with five layers of \BN{} is shown in figure \ref{fig-p3}.

Dipoles are clearly visible at the interfaces between the different materials. The largest interface dipole is between Cu and \BN{}. Electrons are piled up on the Cu surface implying depletion close to the \BN{} surface. A similar effect is observed in the physisorption of organic molecules on metal surfaces,\cite{Rusu:jpcc09,Rusu:prb10} or even in the adsorption of noble gas atoms on metal surfaces.\cite{Bagus:prl02} There it is attributed to Pauli exchange repulsion between the adsorbate and the substrate, which results in a ``push-back'' of electrons towards the ``softer'' material, in this case the metal substrate.\cite{Bagus:prl02} This mechanism for interface dipole formation is quite general, and we speculate that the dipole at the  Cu|\BN{} interface has a similar origin. Between \BN{} and graphene, one observes a similar but significantly smaller interface dipole. These dipoles are truly localized at the interfaces; their sizes do not depend on the number of \BN{} layers as long as there is more than one. Both dipoles can also be obtained in separate calculations for the two interfaces, i.e. one for \BN{} adsorbed on Cu(111), and one for graphene adsorbed on \BN{}.

An interface dipole layer results in a discontinuity in the potential energy perpendicular to the interface. The potential energy step at an A$|$B interface can be obtained from an A$|$B slab calculation as the difference between the work functions on the A and the B sides of the slab. In practice, a single \BN{} layer on top of the Cu(111) surface is sufficient to calculate the potential energy step $\Delta_{\mathrm{Cu}|\mathrm{BN}}$ at the Cu(111)$|$\BN{} interface. Similarly, the potential energy step $\Delta_{\mathrm{BN}|\mathrm{C}}$ at the \BN{}$|$graphene interface can be determined with a system comprising two \BN{} layers and the graphene sheet.

We find $\Delta_{\mathrm{Cu}|\mathrm{BN}}=1.12$ eV, and $\Delta_{\mathrm{BN}|\mathrm{C}}=0.14$ eV. Without additional charge transfer the difference between the Fermi levels in Cu and graphene in the Cu(111)$|$\BN{}$|$graphene structure would be 
\begin{equation}
V_0=W_\mathrm{M}-W_\mathrm{G}-\Delta_{\mathrm{Cu}|\mathrm{BN}}-\Delta_{\mathrm{BN}|\mathrm{C}}
\label{V0}
\end{equation}
which we calculate to be $V_0=-0.49$ eV. To achieve equilibrium requires transferring electrons between Cu and graphene to set up an electrostatic potential that compensates for $V_0$. The sign of $V_0$ indicates that electrons are transferred from Cu to graphene, which results in $n$-type doping, i.e. a negative value for $\Delta E_{\rm F}$, in agreement with figure \ref{A}.

\textit{External Field.} The charge transfer leads to an intrinsic electric field across the \BN{} slab, which polarizes the \BN{} layers. This polarization is clearly identifiable in figure \ref{fig-p3} as small oscillations of $\Delta n(z)$ in the \BN{} region in the absence of an external electric field (red line). It can be eliminated by applying an external electric field that opposes the intrinsic field. With $E_\mathrm{ext}=+0.1$ V/\AA\, the total internal electric field is zero, and the charge distribution in the \BN{} slab becomes identical to that of a free-standing \BN{} slab (black line). The blue line in figure \ref{fig-p3} shows $\Delta n(z)$ resulting from reversing the external field $E_\mathrm{ext}=-0.1$ V/\AA, that increases the  polarization of the \BN{} slab.

\begin{figure} 
\includegraphics[scale=0.33, bb= 30 21 721 480]{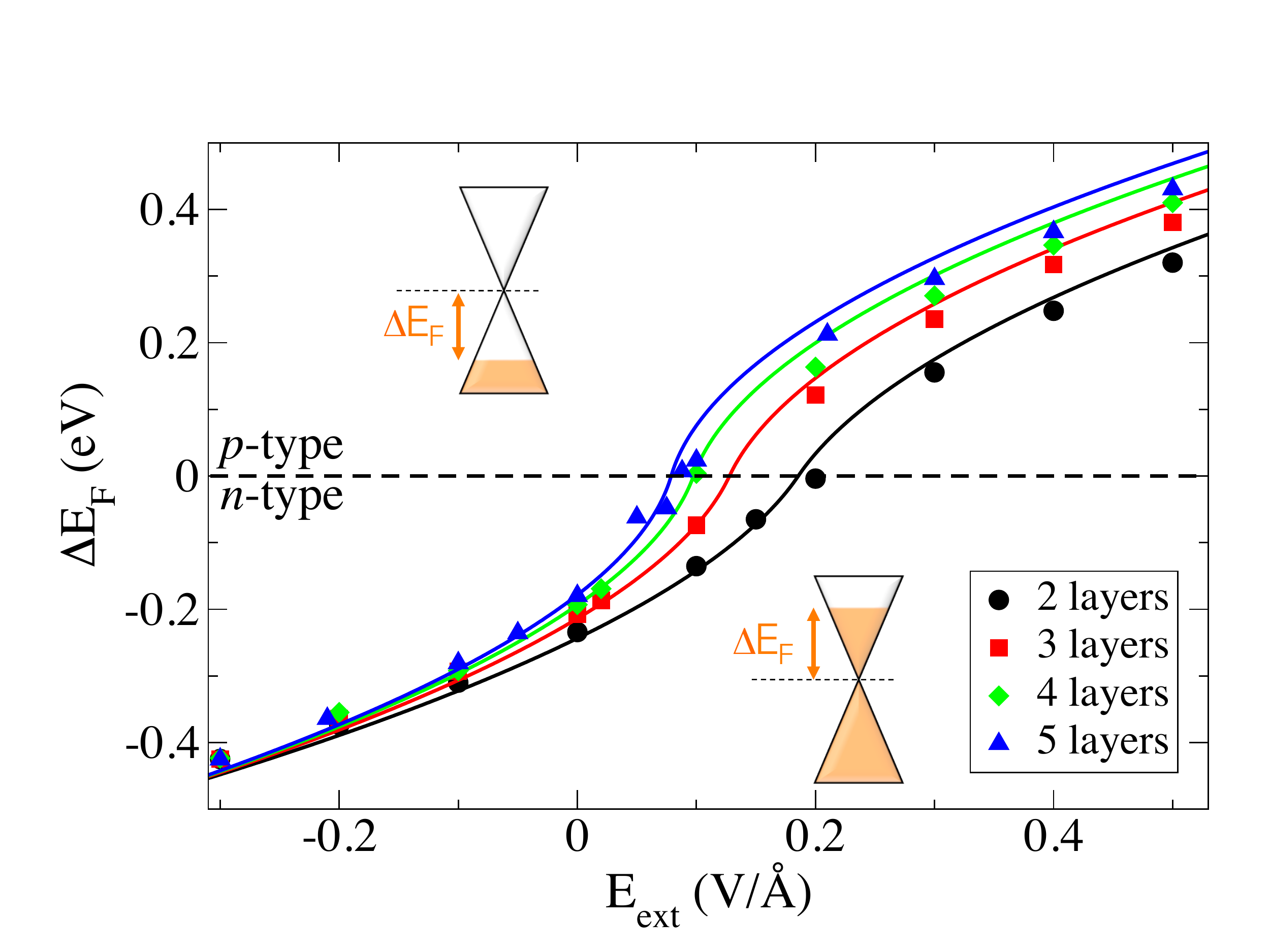}
\caption{Fermi level shift as a function of an external electric field for Cu(111)$|$\BN{}$|$graphene structures with 2-5 layers of \BN{}. The symbols indicate calculated DFT values; the lines represent the model of eq. \ref{fermishiftE}.} \label{B}
\end{figure}

An external field can also be used to control the position of the Fermi level, i.e. the concentration of charge carriers, in graphene. \cite{Novoselov:nat05,Zhang:nat05} Calculated Fermi level shifts for a range of \BN{} slab thicknesses and external electric field values are plotted in figure \ref{B}. The curves for different \BN{} slab thicknesses have a similar, highly nonlinear, shape. Similar shapes of the position of the Fermi level as a function of a gate voltage have been observed in scanning tunneling spectroscopy (STS) experiments,\cite{Zhang:natp08,Zhang:natp09,Xue:natm11,Decker:nanol11} as well as in work function measurements.\cite{Yu:nanol09} 

The points on the curves where $\Delta E_{\rm F}=0$ correspond to the charge neutrality level of graphene, i.e. to undoped graphene. At these points the external field $E_\mathrm{ext}$ is equivalent to a potential difference $V_\mathrm{g}$ across the \BN{} slab that compensates for $V_0$. We expect $|E_\mathrm{ext}| \propto |V_\mathrm{g}|/d$, where $d \sim D$ the thickness of the \BN{} layer. The external field strength corresponding to the charge neutrality level should then decrease monotonically with increasing slab thickness, as is indeed observed in figure \ref{B}.

\begin{figure}
\includegraphics[scale=0.33, bb=30 39 770 520]{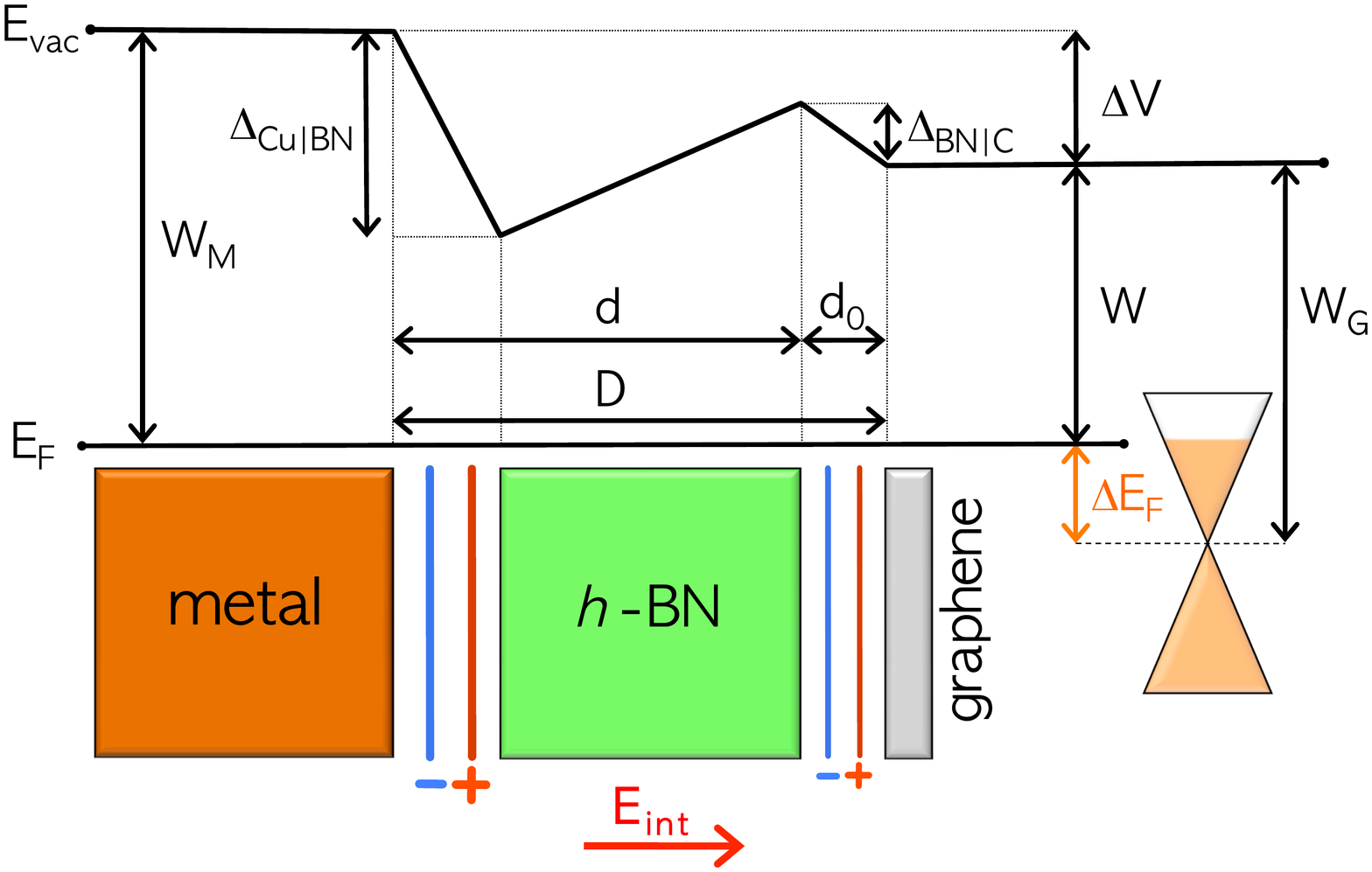}
\caption{Schematic drawing of the metal$|$\BN{}$|$graphene structure and the electrostatic potential (energy) across the structure. $E_{\rm int}$
represents the total electric field across the \BN{} slab of thickness $d$; the other symbols are explained in the text.} \label{fig:potential}
\end{figure}

\textit{Model.} To understand the intrinsic doping (fig. \ref{A}), as well as the external field effect (fig. \ref{B}) quantitatively, we develop the following analytical model whose parameters are shown in figure \ref{fig:potential}. The work function of the metal$|$dielectric$|$graphene stack is given by $W = W_{\rm M} - \Delta V$, with $\Delta V$ the total potential difference across the stack. We model this potential difference as $\Delta V(d) = -e E_{\rm int} d + \Delta_{\mathrm{Cu}|\mathrm{BN}} + \Delta_{\mathrm{BN}|\mathrm{C}}$, where $E_{\rm int}$ is the total electric field inside the dielectric, and $e>0$ is the elementary charge. This field can be related to an externally applied electric field by $\epsilon_0 (E_{\rm ext} - \kappa E_{\rm int}) = \sigma$, with $\kappa$ the dielectric constant of the dielectric layer and $\sigma$ the surface charge density of the graphene sheet. These relations can be used in eq. \ref{eq1} to derive a first expression for $\Delta E_{\rm F}$ in terms of $\sigma$.

A second relation between $\Delta E_{\rm F}$ and $\sigma$ is obtained by noting that charge in graphene is introduced by (de)populating states away from the charge neutrality point, $\sigma=e\int_0^{\Delta E_{\rm F}} D(E)dE$, and that the density of states near the conical points is well described by a linear function, $D(E)=D_0|E|/A$, with $D_0=0.09$/(eV$^2$ unit cell)\cite{Giovannetti:prl08} and $A=5.18$ \AA$^2$ the area of a graphene unit cell. This then gives $\sigma = {\rm sign}(\Delta E_{\rm F})\Delta E_{\rm F}^{2}\, eD_{0} /(2A)$. Combining the two relations between $\Delta E_{\rm F}$ and $\sigma$ gives
\begin{equation}
\Delta E_{\rm F}=\pm\frac{\sqrt{1 + 2\alpha D_0\, d\left\vert V_{\rm g} - V_0 \right\vert/\kappa}  - 1}{\alpha D_0\, d/\kappa},
\label{fermishiftE} 
\end{equation}
where $V_{\rm g} = -e E_{\rm ext}d/\kappa$, $V_0$ is given by eq. \ref{V0}, and $\alpha= e^2/\epsilon_0 A = 34.93$ eV/\AA. The sign of $\Delta E_{\rm F}$ is determined by the sign of $V_{0} - V_{\rm g}$. Equation \ref{fermishiftE} describes how the Fermi level in graphene depends on the gate voltage $V_{\rm g}$ (or the external electric field $E_{\rm ext}$) and the thickness $d$ of the dielectric layer. We define $d$ as the separation $D$ between the top Cu layer and the graphene sheet minus a correction $d_0$, because most of the displaced charge is localized between Cu and \BN{} and between \BN{} and graphene, see figure \ref{fig-p3}. We have used the value $d_0 = 2.4$ \AA.\cite{Khomyakov:prb09}  In principle the dielectric constant of the \BN{} layer $\kappa$ is a (weak) function of its thickness. We have used the constant value $\kappa = 2.72$, which we calculate for a two-layer \BN{} slab by putting the isolated slab in an external electric field\cite{Resta:prb86}. We then determine the internal field by calculating the macroscopic average $\bar{V}(z)$\cite{Baldereschi:prl88} of the electrostatic potential and differentiating it. The dielectric constant is then defined as the ratio between the internal and external field.

The results of this model are given in figures \ref{A} and \ref{B}. The agreement with the results from the DFT calculations for the Cu(111)$|$\BN{}$|$graphene structures is very good. Note that the model has no adjustable parameters, as quantities such as $V_0$ (eq. \ref{V0}) are obtained from separate calculations. The model correctly describes the intrinsic doping and its dependence on the \BN{} layer thickness, fig. \ref{A}, as well as the dependence of the doping on the external field, fig. \ref{B}.

\textit{Summary.} The doping of graphene in Cu(111)$|$\BN{}$|$graphene structures is studied by monitoring the Fermi level shift by means of first-principles DFT calculations. We predict that graphene is intrinsically $n$-type doped for sufficiently thin \BN{} layers. This is due to the potential difference between Cu and graphene arising from substantial dipole layers at the Cu$|$BN and the BN$|$graphene interfaces, as well as from the work function difference between Cu and graphene, eq. \ref{V0}. The doping level decreases with increasing \BN{} layer thickness, and approaches zero for thick layers. It can be varied by applying an external electric field and the resulting shift of the Fermi level has a modified square-root like dependence on the field. For thick dielectric spacers, and in the absence of work-function-difference and interface-dipole terms, this is similar to what is observed in experiments.\cite{Zhang:natp08,Zhang:natp09,Xue:natm11,Decker:nanol11,Yu:nanol09} For very thin dielectric layers, these interface terms are predicted to play an important role and both the \BN{} layer thickness dependence as well as the field dependence of the doping can be described quantitatively by an analytical model, eq. \ref{fermishiftE}. The parameters of this model can be determined experimentally or obtained from DFT calculations on individual surfaces (the work functions of the metal and of graphene), on interfaces (the interface dipoles formed at the Cu(111)$|$\BN{} and the \BN{}$|$graphene interfaces), or on slabs (the dielectric constant of a \BN{} layer). Graphene field-effect devices using a thin layer \BN{} gate dielectric should exhibit an intrinsic doping, and a dependence of the Fermi level on the applied gate voltage that is described by eq. \ref{fermishiftE}.

\textit{Acknowledgement.}We thank Thijs Veening for useful discussions. M.B. acknowledges support from the European project MINOTOR, grant no. FP7-NMP-228424. The use of supercomputer facilities was sponsored by the ``Stichting Nationale Computerfaciliteiten (NCF)'', financially supported by the ``Nederlandse Organisatie voor Wetenschappelijk Onderzoek (NWO)''.

\providecommand*{\mcitethebibliography}{\thebibliography}
\csname @ifundefined\endcsname{endmcitethebibliography}
{\let\endmcitethebibliography\endthebibliography}{}

\end{document}